# Digitally virtualized atoms for acoustic metamaterials

Choonlae Cho[1+], Xinhua Wen[2+], Namkyoo Park[1*], Jensen Li[2*]

[1]Photonic Systems Laboratory, Department of Electrical and Computer Engineering, Seoul National University, Seoul 08826, South Korea
[2]Department of Physics, The Hong Kong University of Science and Technology, Clear Water Bay, Hong Kong, China.

By designing tailor-made resonance modes with structured atoms, metamaterials allow us to obtain constitutive parameters outside their limited range from natural or composite materials. Nonetheless, tuning the constitutive parameters relies much on our capability in modifying the physical structures or media in constructing the metamaterial atoms, posing a fundamental challenge to the range of tunability in many real-time applications. Here, we propose a completely new notion of virtualized metamaterials to lift the traditional boundary inherent to the physical structure of a metamaterial atom. By replacing the resonating physical structure with a designer mathematical convolution kernel with a fast digital signal processing circuit, we show that a decoupled control of the effective bulk modulus and density of the metamaterial is possible on-demand through a software-defined frequency dispersion. Purely noninterfering to the incident wave in the off-mode operation while providing freely reconfigurable amplitude, center frequency, bandwidth, and phase delay of frequency dispersion in on-mode, our approach adds additional dimension to wave moulding and can work as an essential building block for time-varying metamaterials.

+ Equal contribution

* Correspondence: nkpark@snu.ac.kr, jensenli@ust.hk

In the past two decades, metamaterials have revolutionized the way we manipulate classical waves, initially for electromagnetic waves [1-3] and later for acoustic waves [4-11], water waves [12] and recently for elastic waves in solids [13-17]. The ability of metamaterials to obtain physical properties beyond those of natural materials comes from the engineering degrees of freedom in designing artificial structures, split-rings as a representative example, to have resonance with tailor-made properties. Since then, many intriguing phenomena, such as negative refraction and invisibility cloaking, which require the most extreme values of the constitutive parameters were demonstrated [18,19]. These prove that metamaterials can be designed to have a wide range of constitutive parameters and can be inhomogeneous. To further make metamaterials useful in practical situations, tunability or reconfigurability will be required in many applications, ranging from active invisibility cloaks, metamaterial antennas, etc. [20,21]. By optical pumping active materials, mechanically changing geometric parameters using MEMS approach or combining external RLC circuit elements with metamaterial structures, the property of the metamaterial structures can be tuned. The tuning has also been extended to the level of each individual atom when electronics (such as FPGA chip or a computer) at the backend is used to store and change the state of the controlling parameters [18,22-24].

In view of tunability and reconfigurability, active acoustic metamaterials are also attracting increasing attention because the acoustic response can be manipulated through electronically controlled elements and thus can achieve a much wider range of effective parameters [25]. Active acoustic metamaterials have been proved as useful tools to achieve many intriguing physics phenomena, such as PT-symmetric metamaterials [26], sound isolation meta-atoms [27], bianisotropic meta-atoms [28], and topological mechanical metamaterials [29]. Nonetheless, for all types of metamaterials, the tuning depends very much on its actual mechanism in modifying the metamaterial resonance of the physical structures, posing a fundamental challenge to the degree of flexibility and the range of tunability, which is important in many applications requesting real-time tuning. Moreover, it is hard to imagine using common approaches to configure the resonating strength, bandwidth, and phase lag separately because it depends on actual mechanisms of tuning.

Here, we present the concept of virtualization of metamaterials and demonstrate such a concept in manipulating acoustic wave propagation. By replacing the frequency resonating response of a physical metamaterial structure with a mathematically designed frequency dispersion, which is implemented by using a digital convolution in the time domain within a microprocessing unit, the metamaterial structure is virtualized to its digital representation using a software code. Without any physical resonating structure present, our digital representation of the virtualized metamaterial, using a software code allows a very arbitrary specification of the desired resonating frequency response.

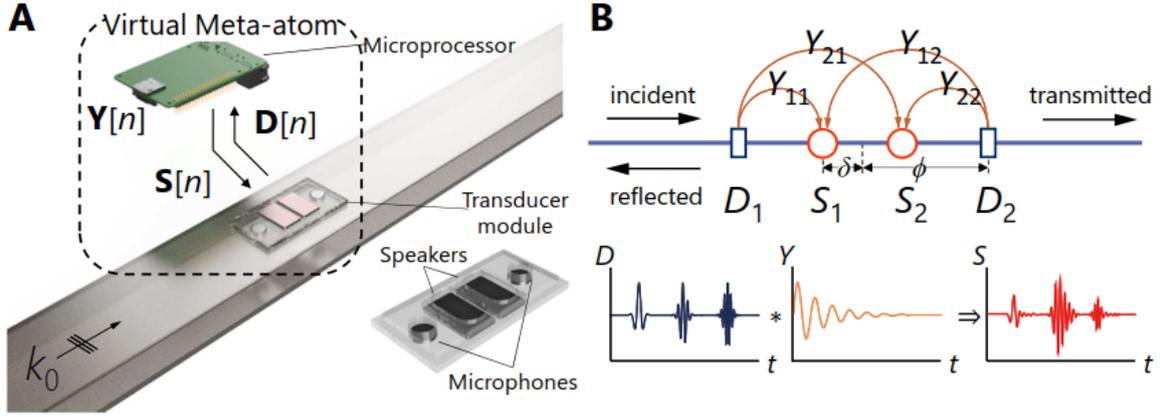

Figure 1 (a) Virtualized metamaterial consisting a structural atom of 2 circular microphones and 2 speakers (the two rectangular patches), connected to a small single-board computer for signal processing at digital level. The virtualized metamaterial is embedded on the inner side of the top cover of a one-dimensional acoustic waveguide, not blocking the incident wave in a passive mode. (b) Schematic representation of the virtualized metamaterial atom: signals detected at the 2 microphones (D1 & D2) are convoluted with a $2 \times 2$ matrix ($Y$) resulting two signals to fire at the 2 speakers (S1 & S2) as secondary radiation from the atom. $Y$ is also called the impulse response of the atom. The phase distance between the two speakers is $2\delta$ (actual distance: $2 \times 8.5 cm$) and the phase distance between the two microphones is $2\phi$ (actual distance: $2 \times 2.6 cm$) for present implementation.

Our virtualized acoustic metamaterial atom is shown in Fig. 1(a). Such an atom consists of a pair of circular microphones, situated outside 2 speakers in the middle. They are bonded on a small rectangular holder to serve as the virtualized atom (lower inset of Fig. 1(a)). The virtualized atom is then put on the inner side of top cover of the one-dimensional hollow waveguide to interact but without blocking the sounds waves travelling within such a waveguide. The microphones and speakers are further connected to an external single-board computer: a Raspberry Pi 3 with an Analog-to-Digital/Digital-to-Analog Conversion module (Waveshare ADS1256 and DAC8532), for operation. Sound waves arriving at the two microphones are detected, digitally sampled, and then real-time processed by a software program running on the single-board computer. The resultant digital signals being output from the program are then converted back to analog signals again and are feedback to the two speakers to generate the synthesized scattered waves. The set of the microphones and speakers with the software program thus defines a generic scattering response of the atom.

The detailed representation of the software program is shown in Fig. 1(b). We construct a general linear operation from the signals at the two microphones D1 & D2 ($D_j(t)$) to the signals at the two speakers S1 & S2 ($S_i(t)$) by operation

$$S_i(t) = -\partial_t^2 \left( \tilde{Y}_{ij}(t - \delta t) * D_j(t) \right), \qquad (1)$$

where $\delta t$ is an extra time delay in the convolution. The whole operation consists of a matrix convolution, a differentiation in time to offset the result of convolution (kernel $\tilde{Y}$ to be designed later) as a driving voltage with zero averaged value for convenience to handle inside the program, and finally a time rate change of the voltage generates the sound radiation by the speaker. In the frequency domain, the operation is summarized as

$$S_i(\omega) = Y_{ij}(\omega)D_j(\omega) \text{ where } Y_{ij}(\omega) = \omega^2 \tilde{Y}_{ij}(\omega)e^{i\omega\delta t}. \tag{2}$$

Each orange arrow in the diagram connects a microphone to a speaker and is then labelled as one of the matrix elements $Y_{ij}$ of the above operation (simply termed as "convolution" from now now). The main horizontal line represents the waveguide direction where an incident wave (e.g. from the left) travels and interacts with the atom. The secondary source at S1 radiates symmetrically in both the forward and backward directions, so as the secondary source at S2. These secondary radiations are added to the incident waves, finally becoming the reflected and transmitted waves within the waveguide. With the $Y_{ij}(\omega)$ specified, it will be possible to solve the overall response of the whole atom in Fig. 1(b), giving transmission/reflection coefficients and the monopolar scattering coefficient $\mathcal{D}_{00}$ (see details in Supplementary Materials for the definition of the scattering matrix $\mathcal{D}$) in terms of $Y$. As the scattering matrix in one-dimensional acoustics is at most $2 \times 2$, we choose to have two microphones and two speakers to detect and generate both monopolar and dipolar incoming and outgoing waves. We also note that all the digital computations in carrying out the convolution can only apply to a finite length of digital signal samples from D1 and D2 before the current sample and has to be finished within one sampling period (133μs) of the Analog-to-Digital Conversion module.

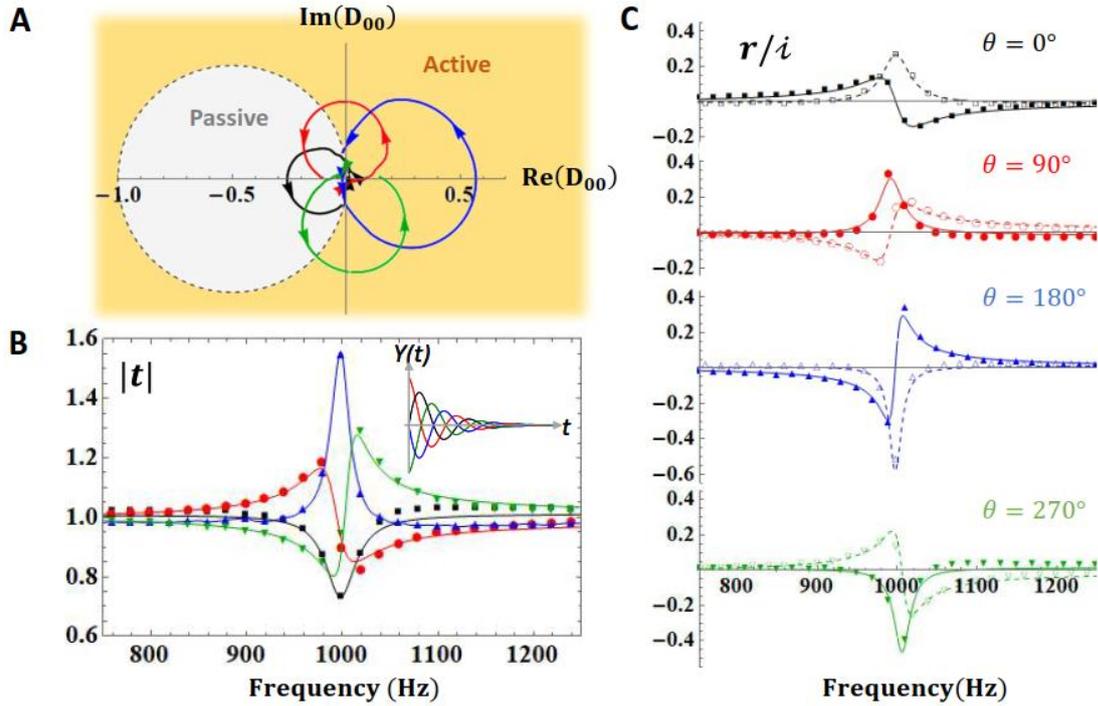

Figure 2 Mimicking Lorentzian frequency dispersion and active acoustic medium with resonating monopolar response. (a) Frequency trajectories of the experimental monopolar scattering coefficient $\mathcal{D}_{00}$ on the complex plane for 4 configurations with convolution phase $\theta = 0°$ (black), 90° (red), 180° (blue) and 270° (green). The yellow/grey area denotes the active/passive region. Arrows indicate the direction from small to large frequencies. (b). Transmission amplitude spectrum for the 4 configurations. The convolution kernel $Y(t)$ for the four different cases of convolution phases are shown in inset. (c) The real part (solid symbols) and the imaginary part (empty symbols) of the complex reflection coefficient $r$. Solid and dashed lines denote the corresponding theoretical Lorentzian line shapes for both the real and imaginary parts respectively.

A Lorentzian frequency dispersion is probably the most representative spectral lines-shape in modelling constitutive parameters (such as permittivity/permeability in electromagnetism and density/modulus in acoustics) from natural materials and metamaterials. It acts like an "alphabet" for both analytical modelling and as a numerical measure to decompose an arbitrary frequency spectrum into a sum of Lorentzian components of different spectral parameters. Here, we would like to instruct our virtualized metamaterial to mimic a Lorentzian response as our first example of virtualized metamaterial. We start by concentrating on monopolar response only for simplicity. It corresponds to an acoustic metamaterial with resonating bulk modulus (the relationship between the effective medium to the $\mathcal{D}_{00}$ and $\mathcal{D}_{11}$ will be described later). For our setting of virtualized atom in Fig. 1, it can be fulfilled by setting $\tilde{Y}_{11} = \tilde{Y}_{12} = \tilde{Y}_{21} = \tilde{Y}_{22} = \tilde{Y}/2$. We consider the convolution kernel $\tilde{Y}(t)$ to have the following form:

$$\tilde{Y}(t) = \frac{a}{\omega_0^2} \sin(\omega_0 t + \theta) e^{-\gamma t} \text{ (for } t > 0\text{) or } 0 \text{ (for } t \leq 0\text{).} \tag{3}$$

It has several model parameters where $\omega_0$ is the resonating frequency, $\gamma$ is the resonating bandwidth, and $a$ is the resonance strength. They are in the unit of radial frequency in the formulas and we specify their values in the unit of frequency by a factor of $1/2\pi$ for brevity. We also define $\theta$ as the convolution phase as an additional parameter to control the shape of frequency dispersion. The software then connects the detector signals to the speaker signals according to Eq. (2) and generates the dimensionless scattering coefficients of the atom as

$$\mathcal{D}_{00}(\omega) = \frac{2\cos\phi\,\cos\delta\,Y(\omega)}{1 - 2e^{i\phi}\,\cos\delta\,Y(\omega)} \cong 2Y(\omega) \tag{4}$$

$$\text{with } Y(\omega) = \omega^2 \tilde{Y}(\omega) e^{i\omega\delta t} = \frac{\omega^2}{\omega_0^2} \frac{a}{2} \left( \frac{e^{i\theta}}{\omega_0 + \omega + i\gamma} + \frac{e^{-i\theta}}{\omega_0 - \omega - i\gamma} \right) e^{i\omega\delta t}.$$

All the other scattering coefficients, $\mathcal{D}_{11}$, $\mathcal{D}_{01}$ and $\mathcal{D}_{10}$, should be zero in this case (see details in Supplementary Materials). For a conventional metamaterial atom, we would expect the monopolar polarizability, being proportional to $\mathcal{D}_{00}/i$, or equivalently one over bulk modulus, to follow a Lorentzian distribution with its imaginary part being positive for a passive atom. For the case $\theta = 0°$, if we choose the convolution delay $\delta t$ so that $\arg(e^{i\omega\delta t}) \cong \pi/2$ is satisfied at resonating frequency $\omega_0$, the resultant $\mathcal{D}_{00}/i$ then mimics the Lorentzian frequency dispersion of a passive acoustic metamaterial in the frequency regime around $\omega_0$. As an example, we choose $\omega_0$ at $1000\,Hz$, $\gamma$ at $15\,Hz$ and a resonating strength $a$ at $7.85\,Hz$ for implementing the case of a passive metamaterial ($\theta = 0°$). $\tilde{Y}(t)$ in Eq. (3) is then programmed as the convolution kernel in the virtualized metamaterial atom. We have experimentally measured the transmission and reflection coefficients in both the forward and backward directions within the waveguide, to calculate the $\mathcal{D}$ matrix. The frequency trajectory of the experimentally extracted $\mathcal{D}_{00}$ from 750 to 1250 $Hz$ is plotted in Fig. 2(a) as the black curve. The trajectory mainly traces out a circle, starting near origin from small frequencies, in the counter clockwise direction. It mainly falls into the passive regime indicated by the grey region with the dashed circle passing through the origin with center at $-0.5$. The complex transmission and reflection coefficients ($t$ and $r$) are simply related to $\mathcal{D}_{00}$ by $t - 1 = r = \mathcal{D}_{00}$. In this case, the resonance causes a dip in the transmission spectrum in Fig. 2(b) while its Lorentzian shape in both the real and imaginary parts is shown clearly when we plot $r/i$ in Fig. 2(c) (the black curves and symbols). The symbols are the experimental results, agreeing well with the theoretical Lorentzian shape in lines. The above constitutes a conventional metamaterial we can adopt with our virtualized metamaterial approach to mimic.

Although Eq. (3) is only a specific class of frequency dispersions, we can now change it by adopting other values of the convolution phase $\theta$ to get a distinctly different virtual atom, most significantly, without the need to design a new physical structure in the conventional approach in designing metamaterials. The software replaces the role of a physical structure. In the case when $\theta$ is changed to 180°, $\tilde{Y}(\omega)$ simply flip signs, so as the sign of the imaginary part of $r/i$. It causes an "anti-Lorentzian" shape of $r/i$ in Fig. 2(c) in blue curves and symbols. The imaginary part now goes negative to indicate a simulated material gain. More intuitively, the transmission coefficient shows up as a peak beyond value one in Fig. 2(b), the additional power going into the transmitted wave is drawn directly from the external digital circuits. Figure 2(a) also shows the trace of $\mathcal{D}_{00}$ for the virtual atom for configurations of different $\theta$, e.g. 90° and 270°. It also traces out circles on the complex plane. In a geometric picture on the complex plane, the convolution phase $\theta$ actually rotates such circles about the origin by the same amount of angle in the clockwise direction. This rotation on the complex plane moves at least part of the circular trajectory out of the passive zone, making the virtual atom unavoidably active. The virtual atom now has the original role of real and imaginary parts of the Lorentzian distribution swapped. The real part of $r/i$ shows up as a peak while the imaginary part shows up as an oscillation instead. The results are shown as red and blue colour in Fig. 2(c). For conventional metamaterials, a Fano resonance is usually introduced to give an asymmetric line-shape [30]. Here, we can create asymmetric line-shape (see the $\theta = 90°/270°$ in the $|t|$ spectrum) by tuning the value of convolution phase.

The virtualized representation of the metamaterial atom (Eq. (3)) provides a straight-forward implementations of an active medium. However, we note that the anti-Lorentzian shape (effectively same as Lorentzian shape but with a negative $a$) has to stand as an approximation in the frequency regime around the resonating frequency. If it is valid for the whole frequency axis, the poles of the complex function $Y(\omega)$ will occur all in the upper half complex plane, denying a causal implementation of the convolution kernel. Our approach guarantees causality as it implements the virtual atom by convolution in the time domain. The approximation on the anti-Lorentzian shape around resonating frequency is linked to the condition $\arg(e^{i\omega\delta t}) \cong \pi/2$ which can only be approximately satisfied around the resonating frequency.

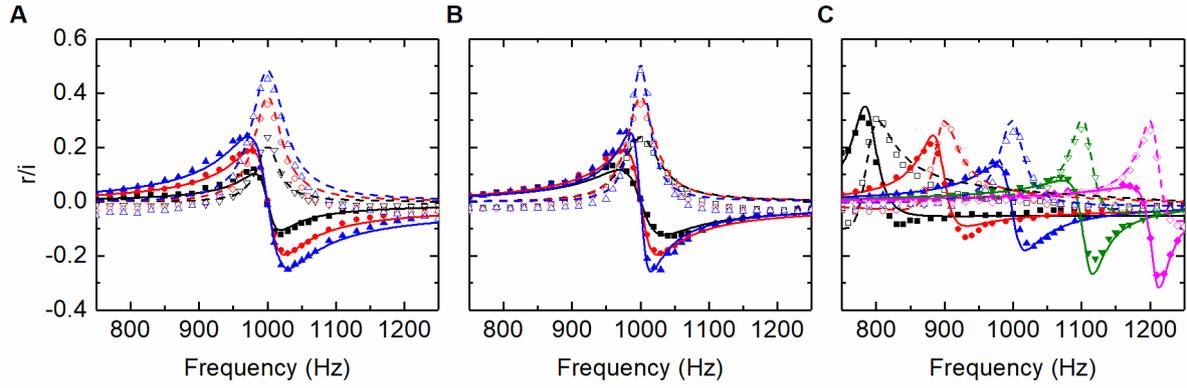

Figure 3 Decoupled tuning of resonance amplitude, bandwidth, and center frequency for virtualized metamaterial. (a) Three cases of resonating strength $a = 3.93$ (black), 7.85 (red) and 11.78$Hz$ (blue) with constant resonating bandwidth $\gamma = 15Hz$ and resonating frequency $\omega_0 = 1kHz$. The real/imaginary part of reflection amplitude $r$ (divided by $\sqrt{-1}$) is plotted in solid/empty symbols for the experimentally results. (b) Three cases of resonating bandwidth $\gamma = 30$ (black), 15 (red) and 7.5$Hz$ (blue) with constant resonating strength $a = 7.85Hz$. (c) Resonating frequency $\omega_0$ varies from 800 to 1200 Hz in steps of 100Hz with $\gamma = 15Hz$ and $a = 7.85Hz$. Here an extra phase shift is inevitable because of the inherent time delay in electronic devices as the resonance frequency increases. The corresponding theoretical models are plotted in solid/dashed lines for the real/imaginary part in all panels.

Our virtualized approach in constructing metamaterial allows us to freely reconfigure the frequency dispersion in a very flexible way on-demand. Conventionally, a physical metamaterial design gives both resonating strength and bandwidth at the same time. These two physical properties (or model parameters) can be reconfigured by two geometric parameters of the metamaterial in principle. However, a decoupled control of the two physical properties by two geometric parameters is highly non-trivial [31]. Varying a single geometric parameter often results to a simultaneous change in both physical properties. On the contrary, as the convolution kernel ($Y(t)$ in Eq. 3) is defined as a mathematical function of time $t$ in the software code for our approach, the resonating strength and bandwidth are simply two input parameters that can be specified independently. Figure 3(a) shows the case for the virtualized atom schematically specified in Fig. 2. The resonating strength $a$ is varied from 3.93, 7.85 to 11.78$Hz$ while the resonating bandwidth is fixed at $\gamma = 15Hz$. The magnitude and spectral profile of both the real part (solid lines/filled symbols) and the imaginary part (dashed lines/empty symbols) of the reflection amplitude increase and scale with $a$. Similarly, we decrease the resonating bandwidth $\gamma$ from 30, to 15 and to 7.5 $Hz$ to obtain sharper resonance with $a$ being kept at a constant value of 7.85$Hz$. The results are shown in Fig. 3(b). In both cases, the experimentally obtained frequency dispersions of reflection amplitude $r$ (plotted in symbols) agree well with theoretical model derived from $Y(t)$ in solid and dashed lines for its real and imaginary part. In fact, as the magnitude of $Y(t)$ decays in time through $\exp(-\gamma t)$, the smallest $\gamma$ we can achieve is limited by the total convolution time ($T_c$) implemented in the software code. A smaller $\gamma$ requires a larger $T_c$ in order to have the magnitude of $Y(t)$ decaying to negligible value before truncation. For example, a request of 10dB decay in $Y(t)$ before truncation is chosen to allow us to implement accurately the target $Y(t)$ with $\gamma$ as small as 3.43 $Hz$. In the mentioned cases, the resonating frequency is kept at 1 $kHz$. Finally, we fix $\gamma = 15 Hz$ and $a = 7.85 Hz$ then vary the resonating frequency $\omega_0$ from 0.8 to 1.2 $kHz$ in steps of 100$Hz$. Clear resonances are observed around the designated resonating frequencies, with a tunable range of resonating frequencies approach almost 40% of the central frequency in the tunable range ($\Delta\omega/\omega$). The tunable range is only limited by the speed of the electronics. Employing faster electronics can have the digital sampling frequency further increased to get a higher frequency bound while the convolution (being accomplished digitally within one sampling period) can involve more number of samples with faster electronics, i.e. a larger $T_c$ to get a lower frequency bound. We also note that the tunability offered by our approach can become more flexible and generic. Since $Y(t)$ is a mathematical function freely encoded in the software, we can instruct the frequency dispersion to be more general, e.g., to capture multi-resonating frequencies and each of these resonating frequencies can have different strengths, bandwidths and can either be simultaneously passive and active as well.

Connecting monopolar incidence to monopolar scattered waves corresponds to an acoustic metamaterial with a resonating bulk modulus [5,6]. Our virtualized approach can also be used to construct metamaterials with response more general than monopolar scattering. As our atom has enough degrees of freedom in generating both monopolar and dipolar secondary radiations, one can use the same virtualized metamaterial technique to generate a dipolar scattering response, corresponding to an effective resonating density. In this case, we set $\tilde{Y}_{11} = -\tilde{Y}_{12} = -\tilde{Y}_{21} = \tilde{Y}_{22} = \tilde{Y}/2$. In this case, we have the dipolar scattering coefficient given by

$$\mathcal{D}_{11}(\omega) = \frac{2 \sin\phi \sin\delta\, Y(\omega)}{1 + 2ie^{i\phi} \sin\delta\, Y(\omega)} \cong 2 \sin\phi \sin\delta\, Y(\omega). \tag{5}$$

$\tilde{Y}(t)$ and $Y(\omega)$ is still defined in Eq. (3) and the one in Eq. (4) but with subscript "1" added to $a, \gamma$ to indicate the dipolar nature of the model parameters. As demonstration, we set the resonating frequency $\omega_0$ as 1.2 $kHz$, resonating strength $a_1 = 14.25\ Hz$ and linewidth $\gamma_1 = 8\ Hz$. We have also set the convolution phase $\theta = 0°$, corresponding to the passive case. The resultant real and imaginary parts of $\mathcal{D}_{11}/i$ are shown in Fig. 4(a) as the red solid and dashed curves with resonating behaviour. It corresponds to a resonating density (in an effective medium picture of the virtualized metamaterial) with a positive resonating peak in its imaginary part. On the other hand, if we change the convolution phase $\theta$ to 180° (with the same parameters for $\omega_0$, $a_1$ and $\gamma_1$), the resonating atoms is gain-dominating around the resonating frequency, showing up a negative peak in the imaginary part of $\mathcal{D}_{11}/i$ in Fig. 4(b). In the same Fig. 4(a) and (b), we also plot the corresponding values of monopolar $\mathcal{D}_{00}/i$ in black color, which are found to have much smaller amplitudes compared to the instructed dipolar response.

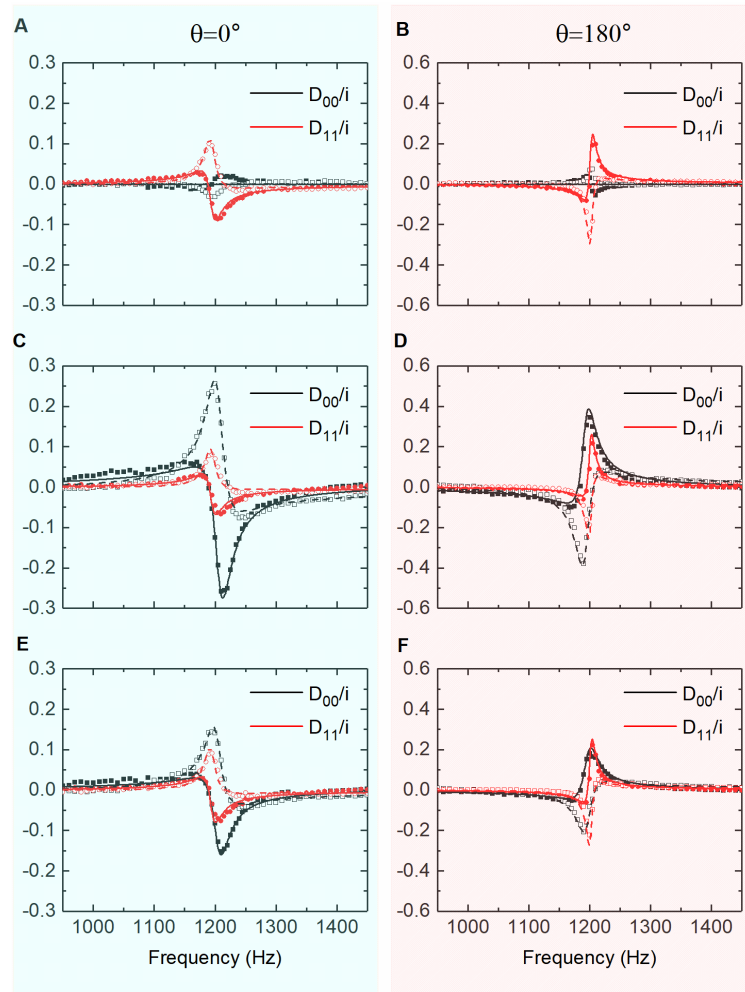

Figure 4 Decoupled control on the monopolar and dipolar scattering coefficients. (a)/(b) Virtualized metamaterial with only dipolar response where the model parameters are set as $\gamma_1 = 8\,Hz$, $a_1 = 14.2\,Hz$, resonating frequency $\omega_0 = 1.2\,kHz$ and convolution phase $\theta = 0°/180°$. Scattering coefficient $\mathcal{D}_{11}/i$ and $\mathcal{D}_{00}/i$ are plotted in red and black color. Solid/dashed lines denote the real/ imaginary parts of the theoretical line shapes. Solid/empty symbols denote the real/imaginary parts of the experimental results. Resonating dipolar $\mathcal{D}_{11}$ dominates over monopolar $\mathcal{D}_{00}$. (c)/(d) Monopolar response is further added to configurations in (a)/(b) with model parameters $\gamma_0 = 15\,Hz$, $a_0 = 6.3\,Hz$ with the same $\omega_0$. (e)/(f) Monopolar response is changed to $a_0 = 4\,Hz$ while other model parameters are kept the same. For all results, the left/right panel shows the scattering coefficients for model parameter $\theta = 0°/180°$.

By taking advantage of the enough degree of freedom of the virtualized metamaterial, the monopolar resonance and dipolar resonance can be generated at the same time. More importantly, all the resonating model parameters can be designed in a very arbitrary manner. For the implementation, we set $\tilde{Y}_{11} = \tilde{Y}_{22} = (\tilde{Y}_0 + \tilde{Y}_1)/2$ and $\tilde{Y}_{12} = \tilde{Y}_{21} = (\tilde{Y}_0 - \tilde{Y}_1)/2$ where $\tilde{Y}_0 / \tilde{Y}_1$ is implemented by Eq. (3) with resonating strength $a_0/a_1$, resonating linewidth $\gamma_0/\gamma_1$ while the resonating frequency commonly set at $\omega_0$ (1.2 $kHz$). Now, both $\mathcal{D}_{00}$ and $\mathcal{D}_{11}$ are resonating, as shown in Fig. 4(c). The detailed model values are given in the caption of Fig. 4. The virtualized atom can also be immediately transferred to the gain regime by changing $\theta$ from 0° to 180°, as shown in Fig. 4(d), in which the resonating peak of the imaginary part for both $\mathcal{D}_{00}/i$ and $\mathcal{D}_{11}/i$ go negative as a dominating gain around resonance. If we represent the system as an effective medium of thickness $d$ (actual thickness for our atom: 6.5$cm$), the relationship between the effective bulk modulus $B$ and effective density $\rho$ can be related to the monopolar and dipolar scattering coefficients through:

$$\chi_0 = B_0/B - 1 \cong \frac{2c_0}{\omega d} \mathcal{D}_{00}/i$$

$$\chi_1 = \rho/\rho_0 - 1 \cong \frac{2c_0}{\omega d} \mathcal{D}_{11}/i$$

where $B_0, \rho_0, c_0$ are the bulk modulus, density and sound speed of the air while $\chi_0$ and $\chi_1$ are called monopolar and dipolar susceptibility here (see Supplementary Materials for the spectra of the effective medium parameters). The capability of controlling both $\mathcal{D}_{00}$ and $\mathcal{D}_{11}$ is essential if we would need to control both transmission and reflection amplitudes, given by $t - 1 = \mathcal{D}_{00} + \mathcal{D}_{11}$ and $r = \mathcal{D}_{00} - \mathcal{D}_{11}$.

For the current virtual implementation of metamaterial, the density and the modulus can be independently tuned without affecting each other, in contrary to conventional metamaterials in which a special kind of atoms has to be designed. Figure 4(e) and (f) are the corresponding results with the same model parameters in Fig. 4(c) and (d) for $\theta = 0°$ and 180° but with the resonating strength $a_0$ divided by a factor of 1.6. From the results, the dipolar resonance is nearly unaffected while the monopolar resonance (e.g. the peak of $Im(\mathcal{D}_{00}/i)$) is divided by around the same factor. Our results show the advantage of the current virtualized approach to design tailor-made configurations that can tackle some of the inherent limitations using conventional metamaterial approaches as the model parameters can now be tuned to have any desired values. This contrasts with common approaches in which resonating strength and bandwidth are unlikely to be independently configurable depending on actual mechanisms to achieve gain. In fact, in conventional metamaterials, it is very often that the dipolar resonance is much sharper than the monopolar resonance. The current approach can make the two resonances have similar shape and bandwidth (Fig. 4(e)). This allows impedance matching (to achieve small reflectance) in a wide frequency regime while the transmission phase can be in resonance. The current virtualized metamaterial approach shows great advantages in modifying the metamaterial resonance.

In conclusion, we have proposed and experimentally demonstrated the notion of virtualized metamaterials, unharnessing the physical restrictions imposed on traditional metamaterials. Directly synthesizing the scattered wave by using a pre-designed kernel function and digitally driven wave sources, it was possible to freely access different dispersion curves on demand, achieving decoupled access on different wave parameters and constitutive parameters. Theoretical background in the conception of the virtualized metamaterial is provided to support intuitive interpretation of the operation, which can be readily generalized to other platforms of wave, such as microwave and electrical circuits. Software-controlled transition between Lorentzian, anti-Lorentzian, and asymmetric dispersion curves are experimentally confirmed within a single platform, at the same time independently addressing the amplitude, center frequency, bandwidth, and convolution phase, for all the dispersion curves and over broad frequency range. The proposed notion of virtualized meta-atom inherently supporting the control of gain/loss in dynamic systems, we expect wide-open applications extending the boundaries of meta-disorder, exceptional points, constant intensity/phase waves, time-varying metamaterials, and topological protections, when combined with the non-Hermitian systems and Dirac/Weyl templates. As well, it should be straightforward to inverse-derive the mathematical kernel on demand for a virtualized metamaterial, for targeted applications and wave parameters. Not limited in acoustic platform, the implementation of kernel function in FPGA also can be envisaged, for faster convolution required in ultra-sonic or microwave applications.

[1] J. Pendry, A. Holten, and W. Stewart, *Magnetism from Conductors and Enhanced Nonlinear Phenomena*. IEEE Trans. Microwave Theory Tech. **47**, 2075(1999).
[2] D. R. Smith, W. J. Padilla, D. C. Vier, S. C. Nemat-Nasser, and S. Schultz, *Composite medium with simultaneously negative permeability and permittivity*. Phys. Rev. Lett. **84**, 4184 (2000).
[3] V. M. Shalaev, *Optical negative-index metamaterials*. Nat. Photon. **1**, 41-48 (2007).
[4] Z. Liu, X. Zhang, Y. Mao, Y. Y. Zhu, Z. Yang, C. T. Chan, and P. Sheng, *Locally resonant sonic materials*. Science **289**, 1734-1736 (2000).
[5] N. Fang, D. Xi, J. Xu, M. Ambati, W. Srituravanich, C. Sun, and X. Zhang, *Ultrasonic metamaterials with negative modulus*. Nat. Mater. **5**, 452 (2006).
[6] J. Li, and C. T. Chan, *Double-negative acoustic metamaterial*. Physical Review E, **70**, 055602(2004).
[7] S. H. Lee, C. M. Park, Y. M. Seo, Z. G. Wang, and C. K. Kim, *Composite acoustic medium with simultaneously negative density and modulus*. Phys. Rev. Lett. **104**, 054301 (2010).
[8] T. Brunet, A. Merlin, B. Mascaro, K. Zimny, J. Leng, O. Poncelet, C. Aristégui, and O. Mondain-Monval, *Soft 3D acoustic metamaterial with negative index*. Nat. Mater. **14**, 384 (2015).
[9] Z. Liang, and J. Li, *Extreme acoustic metamaterial by coiling up space*. Phys. Rev. Lett. **108**, 114301 (2012).
[10] Y. Ding, Z. Liu, C. Qiu, and J. Shi, *Metamaterial with simultaneously negative bulk modulus and mass density*. Phys. Rev. Lett. **99**, 093904 (2007).
[11] S. A. Cummer, J. Christensen, & A. Alù, *Controlling sound with acoustic metamaterials*. Nat. Rev. Mater. **1**, 16001 (2016)

[12] X. Hu, C. T. Chan, Kai-Ming Ho, and Jian Zi. "Negative effective gravity in water waves by periodic resonator arrays." Physical review letters **106**, 174501(2011).

[13] Y. Wu, Y. Lai, and Z. Q. Zhang, *Elastic metamaterials with simultaneously negative effective shear modulus and mass density*. Phys. Rev. Lett. **107**, 105506 (2011).

[14] S. Brûlé, E. H. Javelaud, S. Enoch, and S. Guenneau, *Experiments on Seismic Metamaterials: Molding. Surface Waves*. Phys. Rev. Lett. **112**, 133901 (2014).

[15] P. Wang, F. Casadei, S. Shan, J. C. Weaver, and K. Bertoldi, *Harnessing buckling to design tunable locally resonant acoustic metamaterials*. Phys. Rev. Lett. **113**, 014301 (2014).

[16] R. Zhu, X. N. Liu, G. K. Hu, C. T. Sun, and G. L. Huang, *Negative refraction of elastic waves at the deep-subwavelength scale in a single-phase metamaterial*. Nat. Commun. **5**, 5510 (2014).

[17] Y. Liu, X. Su, and C. T. Sun, *Broadband elastic metamaterial with single negativity by mimicking lattice systems*. J. Mech. Phys. Solids **74**, 158-174 (2015).

[18] Schurig, David, J. J. Mock, B. J. Justice, Steven A. Cummer, John B. Pendry, A. F. Starr, and D. R. Smith. "Metamaterial electromagnetic cloak at microwave frequencies." Science **314**, 977-980 (2006).

[19] S. Zhang, C. Xia, and N. Fang, *Broadband acoustic cloak for ultrasound waves*. Phys. Rev. Lett. **106**, 024301 (2011).

[20] M. Selvanayagam, and G. V. Eleftheriades, *Experimental demonstration of active electromagnetic cloaking*. Phys. Rev. X, **3**, 041011 (2013).

[21] A. K. Sarychev, and G. Tartakovsky, *Magnetic plasmonic metamaterials in actively pumped host medium and plasmonic nanolaser*. Phys. Rev. B, **75**, 085436 (2007).

[22] P. Celli, W. Zhang, and S. Gonella, *Pathway towards programmable wave anisotropy in cellular metamaterials*. Physical Review Applied, **9**, 014014 (2018).

[23] C. Della Giovampaola, and N. Engheta, *Digital metamaterials*. Nat. Mater. **13**, 1115 (2014).

[24] J. Xia, D. Jia, H. Sun, S. Yuan, Y. Ge, Q. Si, and X. Liu, *Programmable Coding Acoustic Topological Insulator*. Adv. Mater. **30**, 1805002 (2018).

[25] B. I. Popa, L. Zigoneanu, and S. A. Cummer, Tunable active acoustic metamaterials. Phys. Rev. B, **88**, 024303 (2013).

[26] R. Fleury, D. Sounas, and A. Alù, *An invisible acoustic sensor based on parity-time symmetry*. Nat. Commun. **6**, 5905 (2015).

[27] B. I. Popa, Y. Zhai, and H. S. Kwon, *Broadband sound barriers with bianisotropic metasurfaces*. Nat. Commun. **9**, 5299 (2018).

[28] Y. Zhai, H. S. Kwon, and B. I. Popa, *Active Willis metamaterials for ultra-compact non-reciprocal linear acoustic devices*. Physical Review B, **99**, 220301(2019).

[29] L. M. Nash, D. Kleckner, A. Read, V. Vitelli, A. M. Turner, and W. T. Irvine, *Topological mechanics of gyroscopic metamaterials*. Proceedings of the National Academy of Sciences, **112**, 14495 (2015).

[30] C. Goffaux, J. Sánchez-Dehesa, A. L. Yeyati, P. Lambin, A. Khelif, J. O. Vasseur, and B. Djafari-Rouhani, *Evidence of Fano-like interference phenomena in locally resonant materials.* Physical review letters, **88**, 225502 (2002).

[31] S. Koo, C. Cho, J.-h. Jeong, and N. Park, *Acoustic omni meta-atom for decoupled access to all octants of a wave parameter space*. Nat. Commun. **7**, 13012 (2016).